\documentclass[12pt]{article}

\usepackage{epsfig}

\newcommand{\be}{\begin{equation}}
\newcommand{\ee}{\end{equation}}
{

\begin{document}
\begin{center}
{\large \bf Berry phase in the simple harmonic oscillator }
\end{center}
\begin{center}
JeongHyeong Park$^\dag$\footnote[4]{E-mail address: 
        jhpark@honam.honam.ac.kr} and
Dae-Yup Song$^\ddag$\footnote[6]{E-mail address: 
        dsong@sunchon.sunchon.ac.kr}
\end{center}
$~~~~~~~~^\dag$Department of Mathematics, Honam University, Kwangju 506-714, Korea\\
$~~~~~~~~~^\ddag$Department of Physics, Sunchon National University, Sunchon 
540-742, Korea

\bigskip
\begin{center}
Short title: Berry's phase \\
Classification numbers: 03.65.Bz  03.65.Ca 03.65.Fd
\end{center}
\bigskip
\begin{abstract}
Berry phase of simple harmonic oscillator 
is considered in a general representation. It is shown that, Berry 
phase which depends on the choice of representation can be defined 
under evolution of the half of period of the classical motions,
as well as under evolution of the period.
The Berry phases do {\em not} depend on the mass or angular frequency of 
the oscillator. The driven harmonic
oscillator is also considered, and the Berry phase is given in terms of 
Fourier coefficients of the external force and parameters which 
determine the representation.
\end{abstract}

\newpage
\section{Introduction}
The (time-dependent) harmonic oscillator gives a system whose quantum
states are described by solutions of the classical equation of motion 
(classical solutions). This fact has been recognized by Lewis 
\cite{Lewis} who found, in an application of asymptotic theory of 
Kruskal \cite{Kruskal},  that there exists a quantum mechanically 
invariant operator. This invariant operator determined by the classical 
solutions has  then been used to construct wave function of (generalized) 
harmonic oscillator systems \cite{Yeon,Ji,Lee}.
An alternative and simple (at least in conceptually) way to 
find the wave functions in terms of classical solutions  is to use 
the Feynman path integral method \cite{Song1}. As observed by Feynman and Hibbs 
\cite{FH}, 
the kernel (propagator) for a general quadratic system is almost determined 
by classical action. The classical action can be given 
in terms of two linearly independent classical solutions and, for a 
general quadratic system, the exact kernel has been found by requiring 
the kernel to satisfy the initial condition and  Schr\"{o}dinger 
equation \cite{Song1}. Further, it has been shown that \cite{Song2}, 
the wave functions of a general quadratic system can be obtained from those 
of unit mass harmonic oscillator system through unitary transformations. 

A perceptive and interesting observation made by Berry \cite{Berry} is that, 
if an eigenstate of Hamiltonian is adiabatically carried around by cyclic 
Hamiltonian, the  change of the phase of wave function separates 
into the obvious dynamical part and an additional geometric part. 
That the geometric change (Berry Phase) still remains some naturalness even 
if the cycled wave function is not an eigenstate of the Hamiltonian and 
even if the carrying is not adiabatic was pointed out by Aharonov and 
Anandan \cite{AA}. In a recent paper by one of the authors \cite{Song3}, 
it was shown that the Berry phase for harmonic oscillator of $\tau$-periodic 
Hamiltonian can be defined only if the two linearly independent
classical solutions are finite all over the time. Moreover, in the cases
of the two linearly independent solutions finite all over the time,
it was shown that there exists at least a representation where the 
Berry phase can be defined under the $\tau$- or $2\tau$-evolution.

The Berry phase is known to be closely related to the 
classical Hannay angle \cite{BH}. By adopting the fact that
a Gaussian wave packet could be a wave function of a harmonic oscillator 
system, the approach of Aharonov and Anandan \cite{AA} has been used to calculate 
Berry's phase, and the relation between 
the phase and Hannay angle is given for the wave packet \cite{GC}.

For the simple harmonic oscillator (SHO) of the equation of motion 
\be
M(\ddot{x} +w^2 x)=0
\ee 
with constant mass $M$ and constant angular frequency $w$, the wave functions become 
stationary if we choose $\cos wt$ and $\sin wt$ as two solutions, while 
in general they describe the states of pulsating probability distribution 
\cite{Song2}.

In this paper, we will calculate Berry phase of the SHO system 
in general representations, and will show that  Berry phase of the SHO system 
indeed depends on the choice of representation. 
We will consider a general representation made from the 
classical solutions $\cos wt$ and $C\sin(wt+\beta)$ of (1), where $C$ is a 
nonzero constant and the constant $\beta$ is not one of 
$(2n+1)\pi/2$ ($n=0, \pm 1, \pm 2,\cdots$). The period of the 
classical solutions is $\tau_0$ (=$2\pi/w$). It will  be shown that,
due to a quasiperiodicity of all classical solution of SHO,
Berry phase can be defined under the $\tau_0/2$-evolution  as well as
$\tau_0$-evolution, which may not be possible for a general oscillator
system considered in \cite{Song3}.
The Berry phase  will be evaluated in terms of the parameters $C$ and $\beta$;  
it turns out that the Berry phase does not depend on the mass $M$ or angular 
frequency $w$ of the oscillator.
For the presence of external force, as already noted in \cite{Moore,Lee},
the phase can not be defined if a particular solution of the equation of motion
diverges as the time goes to infinity. For the cases where the particular 
solution is periodic, the Berry phase whose leading order is order of $1/\hbar$ 
will be evaluated in terms of the Fourier coefficients of the external force.
In view of view analyses on general harmonic oscillator of \cite{Song3}, 
the SHO is rather special in the sense that any number can be the period of the 
Hamiltonian. As will be mentioned in detail, we thus need care in applying the 
formulas to find Berry phase of the SHO system.

\section{The wave functions and (quasi)periodicity}
As is well-known \cite{Song2,KL,Yeon,Ji,Lee},  the wave functions of 
SHO can be written as
\begin{eqnarray}
\psi_n(x,t)
&=& 
     {1\over \sqrt{2^n n!}}({\Omega \over \pi\hbar})^{1\over 4}
     {1\over \sqrt{\rho(t)}}[{u(t)-iv(t) \over \rho(t)}]^{n+{1\over 2}}
     \exp[{x^2\over 2\hbar}(-{\Omega \over \rho^2(t)}
               +i M(t){\dot{\rho}(t) \over \rho(t)})]
\cr & &~~~~~~~~
        \times  H_n(\sqrt{\Omega \over \hbar} {x \over \rho(t)}),
\end{eqnarray}
where the $u(t),v(t)$ are two linearly independent solutions of (1) and $\rho(t)$ is 
defined as
\be
\rho(t)= \sqrt{u^2(t)+v^2(t)}.
\ee 
Without losing generality, the two linearly independent solutions  $u(t),v(t)$ can be
written as
\be
u(t)= \cos wt ~~~~~~v(t)=C\sin (wt+ \beta).
\ee

If we choose $C=1$ and  $\beta=0$ , it gives the "stationary representation" where
the wave functions are of stationary probability distribution
\[
\tilde\psi_n(x,t)= {1\over \sqrt{2^n n!}}({Mw \over \pi\hbar})^{1/4}
\exp[-i(n+{1\over 2})wt -{Mw\over \hbar} x^2]H_n(\sqrt{Mw\over \hbar} x).
\]
The $\tilde\psi_n(x,t)$ is the eigenstate of the Hamiltonian of SHO
\be
H={p^2\over 2M}+{Mw^2 \over 2}x^2,
\ee
with eigenvalue $E_n=(n+{1\over 2})\hbar w$.
The Hamiltonian does not depend on time, so the Hamiltonian is periodic 
with any period. 
The analyses in \cite{Song3} might thus suggest that, in the stationary 
representation  where the Berry phase can be defined for an evolution of any time, 
the phase is zero.  Indeed, phase change of $\tilde\psi_n$  over $t$ evolution 
is simply  equals to the dynamical phase change $-E_n t/\hbar$, so that 
Berry phase is zero.

\begin{figure}
\begin{center}
\epsfig{file=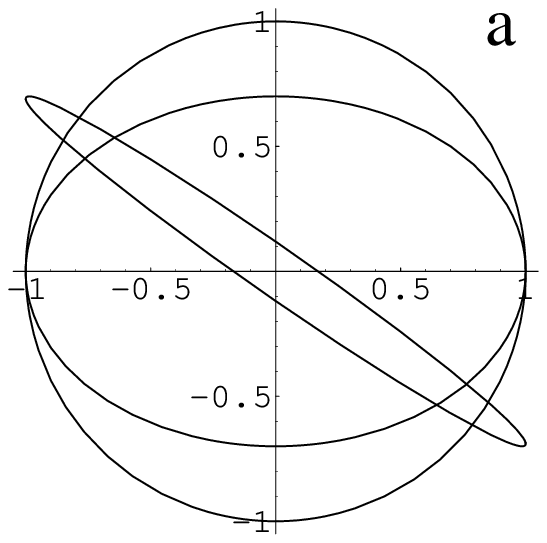, height=5.5cm, width=5.5cm}
$~~~~~~~~~$
\epsfig{file=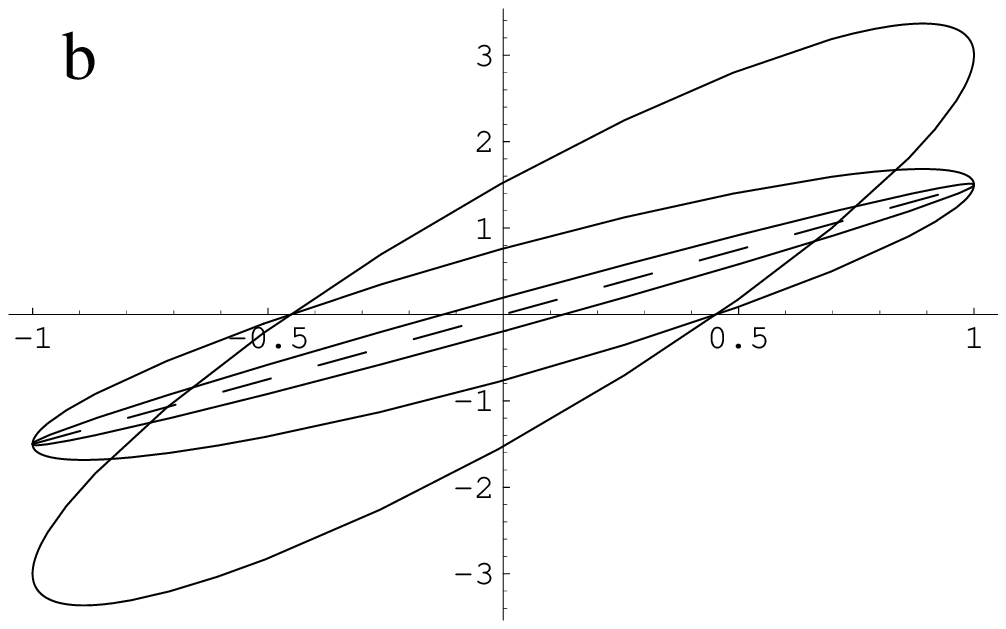, height=5.5cm, width=7.0cm}
\end{center}
\caption{\small The representation space of SHO: the horizontal and vertical axes
denote $u$ and $v$, respectively. Some of the curves which would depict
classical motions of the SHO are given. Though the classical motion for the
dashed line may be allowed, representation corresponding to the line does not exist.}
\bigskip
\end{figure}

In the figure 1, some trajectories of $(u(t),v(t))$ which would depict
classical motions of SHO are given. They make closed curves, since both of $u(t)$ and 
$v(t)$ are periodic.    
Different closed curves in the figure 1 give different sets of wave functions, 
while it is not possible to construct a representation corresponding to the dashed line 
in {\bf b} of the figure. 
Classical motion of SHO may be depicted as the mass $M$ circulating along 
a closed curve with uniform angular velocity $w$, so that it needs a period 
$\tau_0$ for a complete circulation. 
As noted in \cite{Ji,Lee,Song3}, if $\rho(t)$ is periodic with some period, 
the wave functions are (quasi)periodic with the period of $\rho(t)$. 
Since $\rho(t)$ is the distance from the origin to a point of a curve, 
as in the figure 1, it is clear that in general 
$\rho(t)$ is periodic with the half of the period of classical motion.
The circle of radius 1 in {\bf a} gives rise to the stationary representation.  
$\rho(t)$ is constant along the circle and thus any number can be
a period of $\rho$, which is compatible with that Berry phase in the stationary 
representation must be 0. 

From the continuity, one may find the relation
\be
u(\tau_0/2+t)-iv(\tau_0/2+t)= \exp(-i\pi)[u(t)-iv(t)],
\ee
which gives the quasiperiodicity relation of the wave functions
\be
\psi_n(x, \tau_0/2+t)= \exp[i\chi_n(\tau_0/2)]\psi_n(x,t)
\ee
with the over all phase change 
\be
\chi_n(\tau_0/2)=-(1/2+n)\pi.
\ee
Since a phase is defined only up to an additive constant $2\pi$, (8) can be written 
in different ways. For example, a relation $\chi_n(\tau_0/2)=(n-1/2)\pi$ is 
equivalent to (8).  

\section{Berry Phase}
The Berry phase $\gamma_n$ is given from the overall phase change by subtracting  the 
dynamical phase change
\be
\gamma_n(\tau_0/2)=\chi_n(\tau_0/2) - \delta_n(\tau_0/2),
\ee
and the dynamical phase change $\delta_n$ is given as \cite{Song3}
\begin{eqnarray}
\delta_n(\tau')&=&-{i\over \hbar}\int_0^{\tau'} 
                  \int_{-\infty}^{\infty}\psi_n^*(t) H \psi_n(t) dx dt \cr
&=&-{1 \over 2}(n+{1\over 2})\int_0^{\tau'}
         [  {\Omega \over M\rho^2}
                          + {M\dot{\rho}^2 \over \Omega}
              +{\rho^2(t)Mw^2  \over \Omega}] dt. 
\end{eqnarray}
For the classical solutions of (4), $\Omega$ defined as $M(u\dot{v}-v\dot{u})$ is 
\be
\Omega= MCw\cos\beta,
\ee
and thus the $\delta_n(\tau_0/2)$ is written as 
\be
\delta_n(\tau_0/2)= -(n+{1\over 2})\pi {1+C^2 \over 2C\cos \beta}.
\ee
The Berry phase of the wave function $\psi_n$ for the $ \tau_0/2$-evolution, therefore,  
can be given as 
\be
\gamma_n(\tau_0/2)=(n+{1\over 2})\pi[-1+{1+C^2 \over 2C\cos\beta}].
\ee
The Berry phase under $\tau_0$-evolution can be obtained from $\gamma_n(\tau_0/2)$ 
\be
\gamma_n(\tau_0)=2\gamma_n(\tau_0/2)= 
           (n+{1\over 2})\pi[{1-2C\cos\beta +C^2 \over C\cos\beta}].
\ee
The $\gamma_n$ depends on $C$ and $\beta$, but does {\em not} depend on the parameters 
of the Hamiltonian $M$ and $w$.  

The fact that the phase $\gamma_n$ is defined up to an additive constant $\pm2\pi$
says that there are infinitely many representation which give
the same Berry phase. For example, if $\cos \beta = 1/2$, the values of $C$ which gives
$\gamma_n(\tau_0/2)=0$ can be written as
\[
(\cdots,  9/2\pm \sqrt{(9/2)^2 -1},5/2\pm \sqrt{(5/2)^2 -1},
 -3/2\pm \sqrt{(3/2)^2 -1},-7/2\pm \sqrt{(7/2)^2 -1}, \cdots).
\]

In order to compare with the the result of Ge and Child \cite{GC},  by defining
\[
\alpha(t)={1\over 2\hbar}({\Omega \over \rho^2} -i M {\dot{\rho} \over \rho}),
\]
one may find the relation
\be
-{i\over 2}\int_0^{\tau_0} {\dot{\alpha} \over \alpha+\alpha^*} dt
    = {1-2C\cos\beta +C^2 \over 2 C\cos\beta}\pi
\ee
which gives the Berry phase $\gamma_0(\tau_0)$ in agreement with the result in (14).

In order to compare the results here with the formulae in \cite{Song3},
we need to consider several cases differently. If we take $\tau$ as one of 
$(2m-1)\tau_0/2$ $(m=1,2,\cdots)$, the Floquet theorem \cite{Magnus} is satisfied 
by that classical solutions are 2$\tau$-periodic. If we take $\tau$ as one of 
$m\tau_0$ $(m=1,2,\cdots)$, the Floquet theorem is satisfied by that classical 
solutions are $\tau$-periodic. In the above cases, the corresponding 
formula for Berry phases \cite{Song3} can be used to obtain the 
Berry phase in (14) or its integral multiples. In a general oscillator, Berry 
phase may not be defined under $\tau/2$-evolution, and thus the equation (13)
has no corresponding formula in \cite{Song3}.
If we take $\tau$ as being not one of $m\tau_0/2$ $(m=1,2,\cdots)$, the fact that
Berry phase can be defined for any evolution in stationary representation
is in agreement with the analyses in \cite{Song3}. 
Again, a formula corresponding to this case can be used to show that 
Berry phase is 0 in the stationary representation.

\section{Driven harmonic oscillator}
For the SHO with a driving force $F(t)$,  the wave function is given as
\cite{Song1,Song2,Song3,Lee}
\begin{eqnarray}
\psi_n^F &=&
     {1\over \sqrt{2^n n!}}({MwC\cos\beta \over \pi\hbar})^{1\over 4}
     {1\over \sqrt{\rho(t)}}[{\cos wt-iC\sin wt \over \rho(t)}]^{n+{1\over 2}}
     \exp[{i\over \hbar}(M\dot{x}_px+\delta(t))]
\cr & &~~~~~~~~
    \times \exp[{(x-x_p)^2\over 2\hbar}(-{\Omega \over \rho^2(t)}
               +i M(t){\dot{\rho}(t) \over \rho(t)})]
          H_n(\sqrt{\Omega \over \hbar} {x -x_p \over \rho(t)}), 
\end{eqnarray}
where 
\be
\delta(t)={M\over 2}\int_{t_0}^t[w^2 x_p^2(z)  -\dot{x}_p^2(z) ] dz
\ee
with arbitrary constant $t_0$. The $x_p$ defined by the relation
\be
\ddot{x}_p + w^2 x_p = {F(t)\over M}
\ee
may be the classical coordinate of $x$. $F$ denotes the driving force, and we only 
consider the periodic $F$ satisfying $F(t+\tau_f)=F(t)$.

If $x_p$ is finite all over the time and there 
exist two positive integers $N,p$ of no common divisor except 1 such that 
$\tau_0/\tau_f=p/N$, the Berry phase can be defined under $N\tau_0$-evolution
in a general representation. The periodic $F(t)$ can be written as 
\be
F(t)=\sum_{n=-\infty}^\infty f_n e^{in w_f t},
\ee
where 
\be
f_n={1\over \tau_f}\int_0^{\tau_f} F(t) e^{-inw_f t} dt
\ee
with $w_f= 2\pi / \tau_f.$
The $x_p(t)$ finite all over the time can be written as 
\be
x_p=\sum_{n=-\infty}^\infty {f_n \over M(-n^2w_f^2 + w^2)}e^{in w_f t}
    +D e^{i wt} + D^*e^{-i wt}.      
\ee 
If $p=1$, $f_N$ must be zero for the finiteness of $x_p$, which will be assumed 
from now on.
In (21), complex number $D$ is a free parameter and different choice of $D$ gives 
different wave functions of the driven system.
The minimum period of $x_p$ is $N\tau_0$ ($=p\tau_f$) in general so that
\be 
x_p(t+N\tau_0) = x_p(t).
\ee
It has been known that \cite{Song3} the Berry phase of a driven system
separates into the contribution from undriven system and that from the presence of 
driving force. The contribution from the driving force is
written as
\[ {1\over \hbar}\int_0^{\tau'} M\dot{x}_p^2 dt, \]
where $\tau'$ is the period needed for the Berry phase.   
The Berry phase $\gamma_n^F$ 
of driven SHO system under $N\tau_0$ evolution is thus given as
\be
\gamma_n^F(N\tau_0)=
    N\gamma_n(\tau_0)+ {1\over \hbar}\int_0^{N\tau_0} M\dot{x}_p^2 dt.
\ee
After some algebra, from (13) and (20) one may find the relation
\begin{eqnarray} 
\gamma_n^F(N\tau_0)
&=&\pi(n+{1\over 2})N[{1-2C\cos\beta +C^2 \over C\cos\beta}]\cr
 && +2\pi{N^3p^2 \over \hbar Mw^3}
    \sum_{n=-\infty}^\infty  {n^2|f_n|^2 \over (p^2n^2 -N^2)^2} 
               + 2\pi{ MNw \over \hbar} |D|^2.
\end{eqnarray} 
              
There exist the representations where Berry phase can {\em not} 
be defined even with periodic $x_p$. 
For example, in the representation of $D\neq 1$, if $\tau_f/\tau_0$ is an irrational 
number Berry phase can not defined under any evolution of finite time.
However, for the SHO with driving force, there is a special representation
of $C=1, \beta=0$ and $D=0$ where Berry phase can be defined for any periodic $x_p$.
In this representation, Berry phase under $\tau_f$-evolution is written as
\be
\gamma_n^F(\tau_f)= {2\pi w_f \over \hbar M}\sum_{n=-\infty}^\infty
       {n^2|f_n|^2 \over (n^2w_f^2 -w^2)^2}.
\ee 
The presence of this special representation is due to the fact that,
Berry phase can be defined for any evolution in the stationary 
representation of the (undriven) SHO system.

\section{Conclusion}
We have calculated Berry phase of the SHO system in terms of
classical solutions, and have explicitly shown that the phase depends on the choice of 
representations. The Berry phase of undriven system does not depend on the mass
or angular frequency of the SHO, while it depends on two parameters which comes 
from choosing classical solutions. 

For a driven system of periodic $x_p$, the Berry phase, if it exists, separates into the 
contribution from the undriven system and that from the presence of driving force. 
For the SHO system with driving force, 
the contribution from the presence of of driving force which is the order of 
$1/\hbar$ depends on mass, angular frequency, and two more real parameters
coming from the way of choosing classical solutions.
There is a special representation where Berry phase can be 
defined for any periodic $x_p$, while in some representations Berry phase 
can not be defined. For a general harmonic oscillator of time-dependent mass and 
frequency with driving force, it may not be the case that there exists such a special 
representation since stationary representation for the system without
driving force may not exist in general.

While classical solutions of the SHO system is 
$\tau_0$-periodic, the solutions have an additional property 
that absolute value of any classical solution is $\tau_0/2$-periodic.
This fact enables one to define the Berry phase under $\tau_0/2$-evolution as well as 
$\tau_0$-evolution. For a general harmonic oscillator of time-dependent mass and 
frequency, if there exists such an additional periodicity in the absolute value
of all classical solution, making use of the periodicity, it will be possible 
to define the Berry phase for the evolution of shorter period than that of 
the Hamiltonian.  

\bigskip\bigskip
\noindent
{\bf Acknowledgments} \newline
JHP wishes to acknowledge the 
partial financial support of the Korea Research Foundation made in the program, 
year of 1998. DYS acknowledges the partial financial support
from Sunchon National University (Non-Directed Research Fund).


\newpage
\begin{thebibliography}{99} 
\bibitem{Lewis} Lewis H R (1968) {\it J. Math. Phys.} {\bf 9} 1976 \newline
 Lewis H R (1968) 1967 {\it Phys. Rev. Lett.} {\bf 18} 510 
\bibitem{Kruskal} Kruskal R M (1957) {\it Phys. Rev.} {\bf 106} 205
\bibitem{Yeon}Yeon K H,  Lee K K, Um C I, George T F and Pandey L N 1993
   {\it Phys.Rev.} A {\bf 48} 2716 \newline
 Yeon K H, Kim D H, Um C I, George T F and Pandey L N 1997
  {\it Phys.Rev.} A {\bf 55} 4023 
\bibitem{Ji} Ji J-Y, Kim J K, Kim S P and  Soh K-S 1995 {\it Phys. Rev.} A 
   {\bf 52} 3352
\bibitem{Lee}  Lee M-H, Kim H-C and Ji J-Y 1997 {\it J. Korean Phys. Soc.} 
  {\bf 31} 560 \newline
 Kim H-C, Lee M-H, Ji J Y and Kim J K 1996 {\it Phys. Rev.} A 
 {\bf 53} 3767
\bibitem{Song1} Song D-Y 1999 {\it Phys. Rev.} A {\bf 59} 2616
\bibitem{FH} R.P. Feynman and A.R. Hibbs 1965 {\it Quantum Mechanics
   and Path Integrals} (McGraw-Hill Inc: New York) pp 58-60
\bibitem{Song2} Song D-Y 1999 {\it  J. Phys. A: Math. Gen.} {\bf 32} 3449
\bibitem{Berry} Berry M V 1984 {\it Proc. R. Soc. London Ser.} A {\bf 392} 45
\bibitem{AA} Aharonov A and Anandan J 1987 {\it Phys. Rev. Lett.} {\bf 58} 1593 
\bibitem{Song3} Song D-Y 1999 {\it Phys. Rev.} A submitted (= quant-ph/9907062)
\bibitem{BH} Berry M V and Hannay J H 1988 {\it  J. Phys. A: Math. Gen.} {\bf 21} L325
\bibitem{GC} Ge Y C and Child M S 1997 {\it Phys. Rev. Lett.} {\bf 78} 2507 
\bibitem{Moore} Moore D J 1990 {\it J. Phys. A: Math. Gen.} {\bf 23} 5523 \newline
  Moore D J  1991 {\it Phys. Rep.} {\bf 210} 1 
\bibitem{KL} Khandekar D C  and  Lawande S V 1975 {\it J. Math. Phys.} {\bf 16} 384 
\bibitem{Magnus} W. Magnus and S. Winkler, {\it Hill's equation}
           (Dover, New York, 1979) pp4-8

\end{thebibliography}
\end{document}